\def\be{\begin{equation}}
\def\ee{\end{equation}}
\def\bea{\begin{eqnarray}}
\def\eea{\end{eqnarray}}
\def\Eq#1{Eq. \ref{#1}}
\def\Ref#1{Ref. \cite{#1}}
\def\ie{{\it i.e.}}
\def\epsaa{\varepsilon_{a\alpha}}
\def\epsbb{\varepsilon_{b\beta}}
\def\epsba{\varepsilon_{b\alpha}}
\def\epsinaa{\varepsilon_{a\alpha}^*}

\def\eps{\varepsilon}
\def\CI{{\cal I}}
\def\HB{\bar{H}}
\def\om{\omega}
\def\omsq{\omega^2}

\def\omn{\omega^n}
\def\UAa{U^A_a}
\def\UBb{U^B_b}

\def\PAa{P^A_{\alpha}}
\def\PAb{P^A_{\beta}}
\def\PBb{P^B_{\beta}}
\def\Tr{\hbox{Tr}}
\def\CE{{\cal E}}
\def\CF{{\cal F}}
\def\CFB{{\cal F}_b}
\def\rt3{\sqrt{3}}
\def\comm#1{\left[ #1 \right]}
\def\ket#1{| #1 \rangle}
\def\bra#1{\langle #1 |}
\def\kett#1#2{| #1, #2 \rangle}

\documentstyle[aps, pra, graphics, twocolumn, epsfig, floats]{revtex}
\title{Mutually Unbiased Bases and Trinary Operator Sets 
            for N Qutrits}
\author{Jay Lawrence \\ {\it Department of Physics, Dartmouth
          College, Hanover, NH 03755, USA}}
\date{revised \today}
\begin{document}
\maketitle
\begin{abstract}
A compete orthonormal basis of $N$-qutrit unitary operators 
drawn from the Pauli Group consists of the identity and 
$9^N-1$ traceless operators.   The traceless ones partition 
into $3^N+1$ maximally commuting subsets (MCS's) of $3^N-1$ 
operators each, whose joint eigenbases are mutually unbiased.   
We prove that Pauli factor groups of order $3^N$ are isomorphic 
to all MCS's, and show how this result applies in specific 
cases.  For two qutrits, the 80 traceless operators partition 
into 10 MCS's.  We prove that 4 of the corresponding basis 
sets {\it must} be separable, while 6 must be totally entangled 
(and Bell-like).  For three qutrits, 728 operators partition 
into 28 MCS's with less rigid structure allowing for the 
coexistence of separable, partially-entangled, and totally 
entangled (GHZ-like) bases.  However a {\it minimum} of 16 
GHZ-like bases must occur.  Every basis state is described 
by an $N$-digit trinary number consisting of the eigenvalues 
of $N$ observables constructed from the corresponding MCS.
\end{abstract}
\medskip
\centerline{PACS numbers:  03.67.-a, 03.65.Ud, 03.65.Wj}
\section{Introduction}
Systems of three-state particles (qutrits) have been much
under discussion recently because they expand the potential
for quantum information processing and they have been realized 
and controlled experimentally.   Specific realizations for use
in quantum communication protocols include biphotons
\cite{Zhukov,Burlakov}, time-bin entangled photons \cite{Thew},
and photons with orbital angular momentum \cite{Vaziri}.
Qutrit quantum computation with trapped ions has been described 
theoretically\cite{Klimov}, while one-qutrit gates have been 
demonstrated experimentally with deuterons \cite{Das}.
Specific advantages of qutrits over qubits include more secure 
key distributions \cite{BPP,BM,Durt}, the solution of 
the Byzantine agreement problem \cite{Fitzi}, and quantum coin 
flipping \cite{Ambainis}.  

In this paper we describe the general framework for quantum
tomography of $N$-qutrit states provided by mutually
unbiased basis sets \cite{WWF} and the special operators (or
measurements) associated with them.  As $N$ increases, these 
operators retain their fundamental trinary character (having 
three distinct eigenvalues), unlike the more 
familiar operators of compound angular momentum, for example.
This means that the quantum numbers specifying the $N$-qutrit
basis states are $N$-digit trinary numbers:  For separable 
states, each digit corresponds to a statement about a single 
qutrit.  For totally entangled states in which the 
entanglement is shared among all qutrits, every statement 
refers to a joint property of two or more (perhaps all) 
qutrits; no statement refers to a single qutrit.  Such 
descriptions parallel those introduced recently for $N$ qubit 
systems \cite{ZBZ,LBZ}.  In this case, the descriptions rely 
on the existence of $N$ commuting trinary operators.  If 
these are Hermitean, they comprise a ``complete set of 
commuting observables'' in the familiar sense \cite{Gottfried} 
that they define a basis in the Hilbert  space of 
pure states, whose dimension is $3^N$.

But the density matrix $\rho$ describing mixed states resides 
in a vector space of dimension $9^N$, together with all other 
operators on the state space.  This {\it operator} space is 
spanned by a complete and orthonormal set of operators 
(orthonormal in the sense $\Tr(O_i^{\dag}O_j) \sim \delta_{ij}$). 
Appropriate trinary operator ``basis sets'' are easily 
constructed as tensor products of unitary one-qutrit operators 
\cite{Preskill,Bandyo}.  This set of $9^N$ operators 
partitions into $3^N+1$ maximally commuting subsets (MCS's) of 
$3^N-1$ operators each \cite{Bandyo}, expending all operators 
in the basis apart from the identity.  Each of these MCS's 
contains a smaller subset of $N$ operators that provide the
trinary labels for the corresponding basis 
states.  The $3^N+1$ distinct basis sets thus defined are 
{\it mutually unbiased} \cite{Bandyo}, {\it i.e.}, the inner 
product between any two states belonging to {\it different} 
ones has the common magnitude, $3^{-N/2}$.  Since measurements 
within each basis set provide $3^N-1$ independent 
probabilities, the $3^N+1$ basis sets together provide 
$9^N-1$.  This number is just sufficient to determine the 
density matrix (with $\Tr \rho = 1$). 

The partitioning structure described above is guaranteed 
because 3 is a prime number:  \Ref{WWF} proved that if the 
dimension ($d$) of a Hilbert space  is equal to a power of a 
prime number (like $3^N$), then a full complement of $d+1$ 
mutually unbiased basis sets (MUB's) exists.  The number $d+1$ 
is necessary and sufficient to determine a mixed state $\rho$ 
with maximum efficiency \cite{WWF}.  \Ref{Bandyo} proved that 
$d+1$ MUB's exist if and only if a partitioning, complete and 
orthonormal operator set exists, and showed how to construct 
such operator sets from Pauli operators.   These operators 
identify the measurements associated with each MUB.   It is
worth mentioning that while MUB's are desireable for quantum
tomography generally, they have proven instrumental in certain
proposed quantum key distributions \cite{Cerf}, and in the 
solution of the Mean King's Problem for prime power dimensions 
\cite{Aravind}.  The last example is a problem in quantum 
state determination is which both the choice of measurement
(the basis set), and its outcome (the state within the basis
set), are to be determined.

It should be noted in this connection that the partitioning 
property of an operator set is not guaranteed by completeness 
and orthonormality.  A good counterexample in the one-qutrit 
case is the conventional set of generators of SU(3), 
$\lambda_1,...,\lambda_8$ \cite{Joshi}.  This set 
{\it is}~orthonormal 
(\Tr $\lambda_i \lambda_j = 2 \delta_{ij}$), and complete 
(with the inclusion of the identity), but it does {\it not} 
partition into four subsets of two commuting operators each.
So the $\lambda_j$ operators provide a basis for quantum 
tomography \cite{Thew2}, but not one associated with mutually 
unbiased basis sets.  In Sec. II we shall construct Hermitean 
operator basis sets that have two desireable properties:  
They partition so as to define MUB's, and they generate 
$SU(3)$ [or $SU(3^N)$ in the general case].  

We are interested here in the discrete group properties of
the unitary basis operators themselves.  The $N$ operators 
that define basis states also generate, by multiplication, 
all remaining operators in the MCS, plus the identity, thus 
forming a group of order $3^N$.   The full operator basis 
consisting of all MCS's plus the identity (one copy), 
totalling $9^N$ operators, does {\it not} form a group, but 
only because multiplication between operators from different 
MCS's generates one of the three phase factors, 
exp($2 n \pi i/3$).  Thus, a group of order $3 \cdot 9^N$ 
(which we shall call the Pauli Group of $N$ qutrits) consists 
of the operator basis {\it in triplicate}, \ie, each operator 
multiplied by each phase factor \cite{Preskill}.   Its factor 
groups relate closely to the partitioning structure.  Those 
of order $3^N$ are most useful because they are isomorphic 
to all of the MCS groups.  

In the next three sections we discuss the one, two, and three
qutrit cases to illustrate the general principles outlined
above, and also to highlight special properties that emerge 
in each case.   The one-qutrit section describes the Pauli 
Group and illustrates its connection with the (unique) 
partitioning of the eight-operator basis set.  The two-qutrit 
section describes the mutually unbiased bases sets, separable 
and totally entangled, and the corresponding operator sets.  
While many partitions exist, we prove that their {\it 
structure} is unique.  The three qutrit section describes 
three types of mutually unbiased basis sets.  We derive the 
allowed partitioning structures and the coexistence conditions 
for the three types.  The concluding section summarizes the 
results and interprets the trinary descriptions of separable 
and entangled states in terms of complementarity between 
individual and joint properties of many-particle systems.  
Two brief appendices supply background 
material for reference as desired while reading the text.   
Appendix A reviews the connection between mutually unbiased 
bases and the partitioning of an operator set.  Appendix B 
proves a new theorem that describes the general relationship 
between such operator partitionings and Pauli factor groups.  

\section{One Qutrit}

For comparison, recall briefly the Pauli operators for a single 
qu{\it bit}, written in outer product notation \cite{Schw} as
\bea
     Z & = & \ket{n}(-1)^n\bra{n},  \nonumber  \\
     X & = & \ket{n+1}\bra{n},       \nonumber  \\
     Y \equiv XZ & = & \ket{n+1}(-1)^n\bra{n},
\eea 
where summation is implied over the values $n = 0,1$ (for spin 
up and down along $z$), and addition is modulo 2 so that the 
second equation, for example, reads
$X = \ket{1}\bra{0} + \ket{0}\bra{1}$.   The four operators 
$I$, $X$, $Y$, $Z$ form a complete orthonormal set of operators 
on the one-qubit Hilbert space.  The eigenbases of $X$, $Y$, 
and $Z$ are mutually unbiased:  An eigenstate of $Z$ has equal
probabilities of being found in any eigenstate of $X$ or $Y$.
The eight operators $\pm Z$, $\pm X$, $\pm Y$, and $\pm I$ form 
the Pauli Group \cite{Preskill} of a qubit.   A larger Pauli 
Group is sometimes defined as well \cite{Nielsen}, one which 
includes the Hermitean counterpart $iY$ ($Z$ and $X$ being 
hermitean as they stand).  This group consists of the sixteen 
elements ($\pm Z$, $\pm iZ$, ...).   The more exclusive former 
definition conforms to the general one that applies to qutrits.

While the operators defined above are all square roots of the 
identity $I$ ($2 \times 2$), the corresponding operators for 
the qutrit case are all cube roots of $I$ (3 $\times$ 3).  We may
write the complete orthonormal set as $I$, $V$, $X$, $Y$, $Z$, 
$V^2$, $X^2$, $Y^2$, and $Z^2$, since each element has a square,
which is also its inverse.   We may define in analogy with the
above, following Refs. \cite{Preskill,Bandyo,Schw},
\bea
     Z & = & \ket{n}\om^n\bra{n},  \nonumber  \\
     X & = & \ket{n+1}\bra{n},       \nonumber  \\
     Y \equiv XZ & = & \ket{n+1}\om^n\bra{n},  \nonumber  \\
     V \equiv XZ^2 & = & \ket{n+1}\om^{2n}\bra{n}, 
\eea 
where summation is implied over the values $n=0,1,2$, 
(corresponding to spin projections 0, +1, $-$1 respectively), 
addition is modulo 3, and 
\be
      \om = \exp(2\pi i/3).
\ee   
Each of the four operators in Eq. (2) defines, as its eigenbasis, 
one of the four required mutually unbiased basis sets \cite{II2}, 
and each operator, together with its square, forms a maximally 
commuting subset (MCS).     

To facilitate writing the multiplication rules among these 
operators, we introduce a concise notation reflecting their
actions upon the eigenstates of $Z$ (the standard basis):
``diagonal'' ($I$, $Z$, $Z^2$), ``right cyclic'' ($X$, $Y$, $V$), 
and  ``left cyclic'' ($X^2$, $Y^2$, $V^2$).  Then, letting the 
index $l = 0,1,2$ denote the operator within each grouping, 
\bea
   E_l & = & Z^l = \ket{n}\om^{nl}\bra{n},  \nonumber  \\
   R_l & = & XZ^l = \ket{n+1}\om^{nl}\bra{n},  \nonumber  \\
   L_l & = & R_l^{\dag} = \ket{n}\om^{-nl}\bra{n+1}.
\eea
One can also think loosely of $R_l$ as ``raising'' and $L_l$ 
as ``lowering'' operators with respect to the standard basis,  
but the ``cyclic'' designations are more descriptive 
\cite{Byrd}: $R_l$ and $L_l$ are norm-preserving, whereas the 
usual angular momentum raising and lowering operators annihilate 
the uppermost and lowermost rungs of the ladder, respectively.  
The multiplication rules may be constructed immediately.  For 
conciseness, nine expressions are condensed into six by writing 
the last six in the form of commutators:
\bea
   E_lE_m & = & E_{l+m}  \nonumber  \\
   R_lR_m & = & \om^{m-l}~L_{-l-m} \nonumber  \\
   L_lL_m & = & \om^{m-l}~R_{-l-m} \nonumber  \\[.3pc]
  \comm{R_l,E_m} & = & (1 - \om^m)~R_{l+m}  \nonumber  \\
  \comm{L_l,E_m} & = & (\om^{m} - 1)~L_{l-m}  \nonumber  \\ 
  \comm{R_l,L_m} & = & (\om^{m-l} - 1)~E_{l-m}.
\eea
It is apparent from this multiplication table that every one 
of the eight operators, $Z$,...,$V^2$, commutes only with 
itself, its square, and the identity.   Consider for example
$R_l$:  It commutes with $R_m$ only if $m=l$, with $L_m$ only 
if $m=l$ (since $L_m$ is then its square), and with $E_m$ only 
if $m=0$.  It is also apparent that the {\it lack} of 
commutativity resides entirely in the phase factors.  

These phase factors spoil the property of closure under
multiplication and prevent these 9 operators, by themselves,
from forming a group.   But, as mentioned, a group of order 27
(the Pauli Group of a qutrit) {\it is} formed by taking the 
nine basis operators {\it in triplicate}, \ie, each operator 
multiplied by 1, $\om$, and $\omsq$.  

This Pauli Group has a trivial factor group of order 9, whose 
elements consist of the triples 
$\CI = (I, \om I, \omsq I)$, ${\cal Z} = (Z, \om Z, 
\omsq Z)$,..., ${\cal V}^2 = (V^2, \om V^2, \omsq V^2)$.   The 
multiplication ${\cal ZX = Y}$ means that the product of any 
element of ${\cal Z}$ with any element of ${\cal X}$ is equal 
to some element of ${\cal Y}$.   This group is abelian because
no factors of $\omn$ appear in its multiplication table. 

Factor groups of order 3 are more interesting.  One example is 
apparent from Eqs. (5).  The factor group elements, each 
consisting of nine Pauli Group elements, may be denoted by 
\bea
  {\cal E} & = & \omn E_l = ({\cal I, Z, Z}^2) \nonumber \\ 
  {\cal R} & = & \omn R_l = ({\cal X, Y, V}) \nonumber \\ 
  {\cal L} & = & \omn L_l = ({\cal X}^2,{\cal Y}^2,{\cal V}^2),
\label{factor}
\eea
where $n = 0,1,2$, and $l = 0,1,2$.   The multiplication 
table consists of
\be
   {\cal R}^2 = {\cal L}, \hskip.7truecm {\cal L}^2 = {\cal R},   
   \hskip.7truecm {\cal RL} = {\cal LR} = {\cal E}
\ee 
and obvious equalities involving ${\cal E}$.   This group is, 
of course, isomorphic to the group ($1,\om,\omsq$). Equations 
\ref{factor} provide a useful summary of the relationship 
between the 27 Pauli Group elements, the factor group of 
order 9, and the factor group of order 3. 

There is a useful relationship between the last factor group 
and the partitioning structure.   Any of the four MCS's may be 
combined with the 
identity to form a subgroup of the Pauli Group.  In Table I, we
pick ($I,Z,Z^2$).  We define this, in triplicate, as the identity 
element ${\cal E} = ({\cal I,Z,Z}^2)$ of the factor group.  The 
columns below (in triplicate) form the elements ${\cal R}$ and 
${\cal L}$. 

\begin{table}
$$
\vbox{
  \halign{
 # \hfil \quad & \hfil # \hfil \quad & \hfil # \hfil \quad \cr
\noalign{\hrule}
\noalign{\smallskip}
         I & Z & Z$^2$                              \cr
\noalign{\smallskip}
\noalign{\hrule}
\noalign{\smallskip}
           & X & X$^2$                              \cr
	   & Y & Y$^2$                              \cr
	   & V & V$^2$                              \cr
\noalign{\smallskip}
\noalign{\hrule}
\noalign{\medskip}
    ${\cal E}$ & ${\cal R}$ & ${\cal L}$            \cr
\noalign{\smallskip}
\noalign{\hrule}
}}  
$$                            
\caption{The partitioning of 8 one-qutrit operators into 4 
maximally commuting subsets.  If all entries are multiplied by 
1, $\om$ and $\omsq$, then the top line, including $I$, forms 
the identity element ${\cal E}$ of the factor group (\Eq{factor}), 
while the first and second columns (below the line) form the 
${\cal R}$ and ${\cal L}$ elements, respectively.}
\end{table}

The first entry in each column generates all remaining entries 
by multiplication with elements of ${\cal E}$, \ie,
$X {\cal E = R}$ and $X^2 {\cal E = L}$.    There are four such
factorizations corresponding to the four choices of ${\cal E}$.
A second example is
\bea
  {\cal E} & = &  ({\cal I, X, X}^2) \nonumber \\ 
  {\cal R} & = &  ({\cal V, Y}^2, {\cal Z}^2) \nonumber \\ 
  {\cal L} & = &  ({\cal V}^2, {\cal Y, Z}).
\eea
Table I and its four variations illustrate that (i) these 
factor groups are isomorphic to all of the subgroups formed 
by the MCS's, and (ii) every factor group element (apart from 
the identity) contains one and only one operator from every 
remaining MCS.  We prove these two points in Appendix B for the 
general case of $N$ qutrits, and apply them in later sections.

It is instructive to compare the order-three qutrit factor 
groups with their qubit counterparts, which are of order two.  
The identity element in the qubit case could be
${\cal E} = (\pm I, \pm Z)$, with the other element (the 
``flip'' element, say) being ${\cal F} = (\pm X, \pm Y)$.
Permutations of $X, Y$, and $Z$ produce the other factor 
groups of order 2.

Let us turn finally to the observables.  Since the eight 
unitary operators $U = Z,...,V$ and $U^{\dag} = Z^2,...,V^2$ 
form (together with the identity) a complete orthonormal set, 
we can immediately construct an alternative Hermitean basis
through the transformation
\bea
   H = (U-U^{\dag})/i\sqrt{3},  \label{Herm1} \\
   \HB = (U+U^{\dag})/\sqrt{3}.  \label{Herm2}
\eea
The four operators of the type $H$ have the trinary spectrum
($0,\pm 1$), and the four others, $\HB$, have the spectrum
(2,-1,-1)/$\sqrt{3}$.  The relationship between each $H$ and
its compatible partner $\HB$ is
\bea
   \HB = (2I-3H^2)/\sqrt{3}.  \label{Herm3}
\eea
The orthonormality relation among all eight is
\bea
   Tr(H_iH_j) = 2 \delta_{ij},
\eea
where indices run from 1,...8, with (say) odd values 
corresponding to $H$ and even to $\HB$.   The properties of 
orthonormality and tracelessness in fact require that 
compatible {\it Hermitean} partners have different spectra.  
One can then see that the standard generators $\lambda_i$ of 
$SU(3)$ fail to partition, because seven have the spectrum 
($\pm 1,0$) and one has ($1,1,-2)/\rt3$.   The $H_i$ 
operators defined here divide equally among these two
spectra, and it will be clear that the equal division 
generalizes to $N$ qutrits.  The $9^N-1$ partitioning 
Hermitean operators provide an optimally symmetrical set of 
generators of $SU(3^N)$.
 
In the one-qutrit case, the four {\it trinary} $H$-operators  
play a special role:   They define the four mutually
unbiased basis sets completely, they generate their partners 
$\HB$ through \Eq{Herm3}, and they generate all eight 
unitary operators through 
\be
   U = \exp(2\pi i H/3), \hskip.8truecm 
   U^{\dag} = \exp(-2\pi i H/3).
\label{UUdag}
\ee
To verify that Eqs. \ref{UUdag} are equivalent to 
Eqs. \ref{Herm1} and \ref{Herm2}, we expand the exponential 
noting that all odd powers of $H$ are equal to $H$ itself, 
and all even powers are equal to $H^2$:
\be
   U = I + {i\sqrt{3} \over 2} H - {3 \over 2} H^2.
\ee
Equation \ref{Herm3} then shows that $U$ and $U^{\dag}$ are
equal to $(\HB \pm iH)\sqrt{3}/2$, 
which is the inverse of Eqs. \ref{Herm1} and \ref{Herm2}.

\section{Two Qutrits}

A complete orthonormal set of operators on the two-qutrit 
Hilbert space is provided by the 81 tensor products of the
form ($I,Z,...,V^2)_1 \otimes (I,Z,...,V^2)_2$.  Henceforth,
in most expressions  we shall drop both the $\otimes$ symbol 
and the subscripts referring to the individual qutrits, so 
that expressions like $ZX$ or $Y^2V$ or $IZ$ will be 
understood as tensor products.  These operators have the 
following properties in common with their one-qutrit factors:  
Each generates a cyclic subgroup of order 3, namely 
$\big(II,PQ,(PQ)^2\big)$, with $(PQ)^3 = II$, and each has 
the trinary spectrum (1,$\om,\omsq$).  The product of any two 
operators is equal to a single operator multiplied by $\om^n$, 
so that the Pauli Group of two qutrits has order 
$81 \times 3 = 243$.  Each of these properties follows from 
the fact that operators on one qutrit commute with operators 
on the other.  

Clearly there is a factor group of order 81, each of whose 
elements consists of a basis operator in triplicate.  But 
there are more interesting factor groups of order 9, for 
example $({\cal E,R,L}) \otimes ({\cal E,R,L})$, which 
relate to the partitioning of the basis operators.  

\begin{table}
$$
\vbox{
  \halign{
  \hfil # \hfil  & \hfil # \hfil & \hfil # \hfil 
           & \hfil # \hfil & \hfil # \hfil  
	   & \hfil # \hfil & \hfil # \hfil
	   & \hfil # \hfil & \hfil # \hfil   \cr
%\noalign{\hrule}
%\noalign{\smallskip}
\noalign{\hrule}
\noalign{\smallskip}
II & IZ & IZ$^2$ & ZI & ZZ & ZZ$^2$ & Z$^2$I 
& Z$^2$Z & Z$^2$Z$^2$                                   \cr
\noalign{\smallskip}
\noalign{\hrule}
\noalign{\smallskip}
 & IX & IX$^2$ & XI & XX & XX$^2$ & X$^2$I 
& X$^2$X & X$^2$X$^2$                                   \cr
 & IY & IY$^2$ & YI & YY & YY$^2$ & Y$^2$I 
& Y$^2$Y & Y$^2$Y$^2$                                   \cr
 & IV & IV$^2$ & VI & VV & VV$^2$ & V$^2$I 
& V$^2$V & V$^2$V$^2$                                   \cr
 & ZX & Z$^2$X$^2$ & VZ & XY & YV$^2$ & V$^2$Z$^2$ 
& Y$^2$V & X$^2$Y$^2$                                   \cr
 & ZY & Z$^2$Y$^2$ & XZ & YV & VX$^2$ & X$^2$Z$^2$ 
& V$^2$X & Y$^2$V$^2$                                   \cr
 & ZV & Z$^2$V$^2$ & YZ & VX & XY$^2$ & Y$^2$Z$^2$ 
& X$^2$Y & V$^2$X$^2$                                   \cr
 & Z$^2$X & ZX$^2$ & YZ$^2$ & XV & VY$^2$ & $Y^2$Z 
& V$^2$Y & X$^2$V$^2$                                   \cr
 & Z$^2$Y & ZY$^2$ & VZ$^2$ & YX & XV$^2$ & V$^2$Z 
& X$^2$V & Y$^2$X$^2$                                   \cr
 & Z$^2$V & ZV$^2$ & XZ$^2$ & VY & YX$^2$ & X$^2$Z 
& Y$^2$X & V$^2$Y$^2$                                   \cr
\noalign{\smallskip}
\noalign{\hrule}
\noalign{\medskip}
  ${\cal EE}$ & ${\cal ER}$ & ${\cal EL}$ & ${\cal RE}$ 
              & ${\cal RR}$ & ${\cal RL}$ & ${\cal LE}$ 
              & ${\cal LR}$ & ${\cal LL}$               \cr
\noalign{\smallskip}
%\noalign{\hrule}
%\noalign{\smallskip}
\noalign{\hrule}
}}  
$$   
\caption{A partitioning of 80 two-qutrit operators into
10 maximally commuting subsets, each consisting of 8 
elements.   Factor group elements are identified at the 
bottom:  The identity element consists of operators in the 
top row (each ``in triplicate''), and remaining elements 
correspond to the columns directly above them 
(in triplicate).}
\end{table}

The basis operators partition into 10 MCS's of 8 operators 
each.  One such partitioning and the associated MUB's were 
presented in \Ref{Aravind}.  We show another (similar) 
partitioning in Table II together 
with a related factor group.  The factor group is defined 
by choosing its identity element ${\cal EE}$ to be the first 
row in triplicate, {\it i.e.}, ${\cal E = (I,Z,Z}^2)$ for 
both qutrits.  All other elements are identified in the 
bottom row, and these consist of the 9 entries in the 
columns directly above, all in triplicate.  Each such 
element is generated by a multiplication 
such as ${\cal ER} = IX \cdot {\cal EE}$.   While the factor 
group is determined uniquely once its ${\cal EE}$ element is 
specified, there are several ways in which the elements can be 
arranged within each column so that every {\it row} is a 
maximally commuting subset, and consequently, all eigenbases 
are mutually unbiased. However, these special arrangements are 
not likely to occur automatically, because only 12 out of the 
($9 ! )^7$ possible arrangements within this particular 
factorization result in commuting subsets.  The theorem of 
Ref. \cite{Bandyo} proves that a partition exists, and our 
Appendix B proves that it conforms to the structure of a 
factor group as illustrated in Table II.  This structure is 
useful in determining all possible partitions, of which there
are 48.  Before embarking on this task, however, it will be
useful to discuss and classify the basis sets resulting from 
Table II.  
\vskip.5truecm
\centerline{{\bf A. Mutually Unbiased Basis Sets}}
\vskip.35truecm
A ``complete set of commuting'' operators consists of any 
two operators in a given row that are not squares of one 
another.  We shall refer to such operator pairs as 
{\it generators} because, in addition to providing the two 
independent quantum numbers needed to specify the nine basis 
states, they generate the group consisting of the remaining 
operators in their row and the identity.   

We may choose the first and third entries in each row as 
the generators.  Then, because those of the top four rows  
are all one-qutrit operators, the corresponding bases must 
be separable.   The first-row basis states are written 
$\kett{n}{m}_{zz}$, the indices ($n,m$) being two-digit 
trinary numbers that specify the eigenvalues ($\om^n,\om^m$) 
of the one-qutrit factors in the subscript, $Z_1$ and $Z_2$. 
Second-row states are written $\kett{n}{m}_{xx}$, denoting 
products of eigenstates of $X_1$ and $X_2$, and similarly 
in rows three and four.   The full basis {\it sets} are 
denoted by S$(Z,Z)$, S$(X,X)$, and so forth (``S'' for 
separable). 

Generators in all remaining rows ($5-10$) are characterized 
by the failure of commutativity between their individual 
one-qutrit factors.  As in the case of two qu{\it bit}s 
\cite{LBZ}, the joint eigenstates of such operator pairs 
must be totally entangled.  Focusing on the fifth row, we 
look for joint eigenstates of $ZX$ and $VZ$.   Let us 
first write the most general form of eigenstates of $ZX$ 
as expansions in separable states $\kett{n}{m}_{zx}$
belonging to S($Z,X$):
\bea
  \ket{\Psi_0} & = & \big(~a\kett{0}{0} + b\kett{1}{2} +
  c\kett{2}{1}~\big)_{zx}    \nonumber  \\
  \ket{\Psi_1} & = & \big(~d\kett{1}{0} + e\kett{2}{2} +
  f\kett{0}{1}~\big)_{zx}   \nonumber  \\
  \ket{\Psi_{2}} & = & \big(~g\kett{2}{0} + h\kett{0}{2} +
  k\kett{1}{1}~\big)_{zx},   
\eea
where the subscripts of $\Psi$ denote the eigenvalues of
$ZX$ (namely 1, $\om$, and $\omsq$, respectively) and the 
coefficients $a, ..., k$ are arbitrary.   The coefficients
are then fixed by requiring that the general states above 
also be eigenstates of $VZ$.  There are three choices of 
coefficients in each expression, corresponding to the three 
eigenvalues of $VZ$.  The salient feature is that the 
coefficients have equal amplitudes, their relative phases
being integral powers of $\om$.  The physical interpretation
of this is that while $ZX$ takes a definite value, its 
factors $Z$ and $X$ are maximally random.   That is, the sum 
of the two indices (the subscript of $\Psi_n$) is fixed, 
while each index by itself takes all three of its possible 
values with equal probabilities.   With this in mind, we can 
write all nine row-5 basis states in the form
\be
  \kett{n}{m}_{B(zx,vz)} = {1 \over \sqrt{3}} 
  ~\sum_{k=0}^2 C_{nmk} \kett{k}{n-k}_{zx},
\label{row5}
\ee
where $C_{nmk}^3=1$.   The subscript on the left side, 
B$(ZX,VZ)$, refers to the entangled basis {\it set} 
(``B'' for Bell-like).  The  
individual states within this set are denoted by the 
two-digit trinary number ($n,m$) specifying the two 
eigenvalues ($\om^n,\om^m$) of the operators ($ZX,VZ$), 
respectively.   The indices ($n,m$) themselves are just the 
eigenvalues of the {\it trinary Hermitean} counterparts 
defined by \Eq{Herm1}; for example,
\be
  H(ZX) = (ZX - Z^2X^2)/i\sqrt{3}.
\ee
So, restating the physical interpretation in terms of
observables, the eigenvalue ($n$) of $H(ZX)$ is the sum, 
modulo 3, of the eigenvalues of $H(Z_1)$ and $H(X_2)$ ($k$ 
and $n-k$, respectively), the sum being definite while the 
summands are random.  

The randomness applies not just to the one-qutrit factors 
in the two generators as shown above, but to {\it all} 
one-qutrit operators.  This follows most dramatically from 
the stunning appearance of all 16 one-qutrit factors in 
every maximally commuting subset in rows 5 - 10!   So 
{\it any} one-qutrit factor may appear in a generator and 
the above arguments apply.  A concise general proof 
will be given in the following subsection.

Rows 8 - 10 of Table II define states that differ subtly 
from those of rows 5 - 7.  Consider the generators
$Z^2X$ and $YZ^2$ of row 8.  The Hermitean counterpart
of either of these, for example
\be
  H(Z^2X) = (Z^2X - ZX^2)/i\sqrt{3},
\ee
refers to differences, not sums of individual qutrit 
indices:  If joint eigenstates of $Z^2X$ and $YZ^2$ are 
expanded in the same product basis S$(Z,X)$ used for
\Eq{row5}, the result takes the form

\be
  \kett{n}{m}_{B(z^2x,yz^2)} = {1 \over \sqrt{3}} 
  ~\sum_{k=-1}^1 D_{nmk} \kett{k}{n+k}_{zx},
\label{row8}
\ee 
showing that the eigenvalue ($n$) of $H(Z^2X)$ is the 
difference between eigenvalues of $H(X_2)$ and $H(Z_1)$.  
Of course one may choose different operators to label the
states, and in row 8 there is one choice, $XV$, that 
refers to sums, while the other three independent choices
refer to differences.  Rows 8 - 10 share the preference 
for differences, while rows 5 - 7 share the preference for
sums.  These balanced asymmetries can be interchanged but
not removed by renaming the operators, and they appear to 
be inevitable.  

In all the above examples, basis states have been written 
in ``minimal'' form, \ie, three-term expansions for 
entangled states, and single terms (by definition) for 
separable states.   This requires a special choice of 
``quantization axes'' for both qutrits, a choice that is
unique for S states although not for B states.   It is
interesting to note that if one instead expands all MUB
states in the standard basis, S($Z,Z$), then all those
outside of S($Z,Z$) have nine-term expansions, of 
necessity.    A full complement of MUB's was written 
out explicitly in this manner for solving the Mean 
King's problem in 9 dimensions \cite{Aravind}. 
\vskip.5truecm
\centerline{{\bf B. The 4S - 6B Theorem}}
\vskip.35truecm
We now address the problem of enumerating all partitions 
of the operator set, and of proving that all have the same 
structure, producing exactly 4 separable and 6 totally 
entangled basis sets.  We begin by proving that there must 
be exactly 4 separable bases.  Consider any operator 
having an identity factor, say $PI$.   The most general 
MCS to which it may belong is generated by itself and any 
operator that commutes with it, apart from a power of 
itself.  Such an operator must take the form $IQ$, leading 
to the MCS ($II,PI,P^2I) \otimes (II,IQ,IQ^2) - II$, where 
$Q$ may be any of $V, X, Y$, or $Z$, squares being 
redundant.  This subset, or its two generators, define
the separable basis set S$(PQ)$.  Since there are four 
operators of the type $PI$ to begin with, and since each 
must belong to a MCS containing a distinct $IQ$ operator, 
we conclude  1) that there are four MCS's of this type, and 
2) there are 24 distinct choices of these four subsets.  
Correspondingly, there are four separable, mutually 
unbiased basis sets, and 24 distinct choices for this 
quartet.  The distinction is simply one of renaming
the basis states of one qutrit while keeping those of 
the other qutrit fixed.  We may therefore conclude from 
the specific example shown in Table II that the remaining 
six basis sets are totally entangled.  It is instructive, 
however, to prove this result in more general terms.

Since the required separable bases exhaust all operators
containing an identity factor, it suffices to show that
any MCS lacking such operators 
must produce a totally entangled basis set.  We make use 
of \Eq{INVS} of Appendix A, which expresses the projection 
operator $\PAa$ of any state $\alpha$ belonging to the 
basis $A$ in terms of all the corresponding operators,  
$U^A_1,...,U^A_{d-1}$, and the identity $I = U^A_d$. The
coefficients $\epsinaa$ all have unit amplitude, and in 
particular, $\eps_{d \alpha} = 1$.   The overall factor 
of $d^{-1} = 1/9$ guarantees that Tr $\PAa = 1$.  The 
reduced density matrix of either qutrit is the partial 
trace of $\PAa$ over the states of the other.  The 
partial trace vanishes for all operators with 
$a = 1,...,d-1$ because no identity factors appear, 
leaving only $a=d$:
\be
  \rho_1 = \Tr_2 \PAa = {1\over 9} \Tr_2 I = 
  {1 \over 3} I_1,
\ee
where $I_1$ is the identity operator on qutrit 1.
Similarly, $\rho_2 = I_2/3$, showing that the pure state
$\PAa$ is totally entangled.  This proves that all six 
remaining MCS's give rise to totally entangled basis sets.
%As a check, one can also use Eq. (34) to confirm that 
%the reduced density matrices for separable states are 
%in fact projection operators on the one-qutrit spaces. 

It is interesting to note that the absence of identity 
factors in a MCS implies two 
other points encountered earlier: 1) {\it All one-qutrit 
operators must appear as factors.}  To prove this, assume 
that they do not.  Then at least one such factor would 
have to appear twice, producing a pair such as $PQ$ and 
$PR$.  But these  are distinct and commute only if 
$R = Q^2$, in which case their product ($P^2I$) contains 
a single identity factor, producing a contradiction. 
2)  {\it No two generators of such a MCS may have 
compatible one-qutrit factors.}   This follows from 
point (1), noting that $PQ$ and $P^2Q^2$ appear in the
same MCS but cannot be generators.

\vskip.5truecm
\centerline{{\bf C. The final count - 48 partitions}}
\vskip.35truecm
Having counted 24 choices for the four separable basis
sets, we ask finally how many choices remain for 
partitioning the complete operator set. It suffices to 
begin with the specific choices of rows 1 - 4 as written 
in Table II, since all others correspond to relabeling 
the operators on the second qutrit.  To count the 
remaining options for partitioning the operators in rows 
5 - 10, we count the ways to rearrange operators in the 
third column, keeping those of the first column fixed, in 
such a way that these pairs generate compatible rows that 
do not overlap with any other rows.  (This simplification 
rests on the factor group theorem of Appendix B.)  
Consider $ZX$ from the first column.  It commutes with 
$VZ, XZ, YZ$, and no others from the third column.  The 
first and third choices are viable, while the second must 
be ruled out because it generates $YY$, which already 
appears in the third row.   Given either viable choice, 
there remains only one choice for partners of $ZY$ and 
$ZV$ that does not reproduce an operator already existing 
in some row above.  Moving to the last three rows, $Z^2X$ 
commutes with $YZ^2, VZ^2, XZ^2$, and no others from the 
third column.   Given our choice for the fifth row, only 
the first choice is viable:  $VZ^2$ is ruled out because 
it generates $YV$, which already appears in the sixth row, 
and $XZ^2$ is ruled out because it generates $VV$ which 
is found in the fourth row.   Therefore, in total, only 
two choices exist for the last six rows given the first 
four rows.  The choice not shown here appears in Table V 
of \Ref{Aravind}.   With 24 options available for the 
first four rows, there are 48 distinct partitions of the 
full operator set.  Only the last choice, involving just
the totally entangled basis sets, does not correspond to 
a simple relabeling of one-qutrit states.

\section{Three Qutrits}

There are $9^3 = 729$ tensor product operators such as
$VZX^2, Y^2IX, ZII$, etc., that comprise a complete 
orthonormal set.  As before, each generates a cyclic subgroup
of order 3 and has the trinary spectrum.  The product of any 
two is a single operator times $\omn$, so that the Pauli group 
has order $3\times 9^3 = 2187$.  The more interesting factor 
groups are of order 27, for example 
(${\cal E,R,L})^{\otimes 3}$; these relate to the partitioning 
of the basis operator set into $3^3+1 = 28$ maximally commuting 
subsets of $3^3-1 = 26$ operators each.  The new aspect that 
emerges with $N=3$ is that different partitioning 
{\it structures} are possible.   For example, if the number 
of separable basis sets is maximized at 4, then all remaining 
24 basis sets are totally entangled.   However, one may choose
fewer than 4 separable bases, requiring that some others have 
partial entanglement.  Let us begin by illustrating the three
types of basis sets and corresponding MCS's that can occur.   
We shall then determine the partitioning structures that are 
possible, without attempting to count the actual number of 
each type.

Product bases may be denoted by S$(PQR)$, where $P$ is any 
of $V, X, Y$, or $Z$, etc.  These are eigenbases of the 
one-qutrit operators $PII, IQI$, and $IIR$, which generate 
the maximally-commuting subsets
$(I,P,P^2)\otimes(I,Q,Q^2)\otimes(I,R,R^2)$.   As with two
qutrits, there are only four (nonredundant) choices for $P$, 
and therefore at most four such subsets, with four 
corresponding product bases.   As before, of course, there 
are many choices for these quartets.

\begin{table}
$$
\vbox{
  \halign{
 # \hfil \quad & \hfil # \hfil \quad & \hfil # \hfil \quad & 
                                       \hfil # \hfil \quad\cr
\noalign{\hrule}
\noalign{\medskip}
        $ XXX $ & $ YYY $ & $ VVV $ &                     \cr
        $ XYV $ & $ YVX $ & $ VXY $ &                     \cr
	$ XVY $ & $ YXV $ & $ VYX $ &                     \cr
	$ Z^2ZI $ & $ Z^2IZ $ & $ IZ^2Z $ & $ ZZZ $       \cr
\noalign{\medskip}
\noalign{\hrule}
}}  
$$                            
\caption{These thirteen three-qutrit operators and their squares
form a maximally commuting subset whose eigenbasis consists of
totally entangled three-qutrit GHZ states.} 
\end{table}

Totally entangled basis sets in which the entanglement is 
shared equally among all three qutrits (analogous to 
three-qubit GHZ basis sets \cite{LBZ}) arise from MCS's
similar to the example shown in Table III.  (``Similar'' 
means ``same number if $I$ factors.'')   We show half of 
the operators which, together with their squares, comprise 
the full MCS of 26.  Each row, together with its squares, 
forms a subgroup when combined with $ZZZ$, its square and 
the identity.  The three generators of the full MCS may be 
any three operators that do not belong to the same subgroup.  
The operators in the last row are notable because they (and 
their squares) are diagonal in the {\it separable} basis 
S$(ZZZ)$.  The first three operators specify differences 
between two one-qutrit indices, and the fourth operator 
specifies the sum of all three.  A simultaneous eigenstate 
(with the sum and all differences equal to zero, for 
example) is
\be
 \ket{\Psi} = 
 \big(~ a \ket{000} + b \ket{111} + c \ket{222}~\big)_{zzz},
\label{three}
\ee
where the subscript indicates separable states, and the
coefficients $a, b,$ and $c$ are arbitrary.  There are nine 
expressions of the form of \Eq{three} which are simultaneous
eigenstates of fourth-row operators, each being a superposition 
($k = 0,1,2$) of separable states $\ket{k,n+k,l+k}_{zzz}$  with 
arbitrary coefficients.   The fourth-row operators
(or more precisely, their Hermitean counterparts $H$), have 
eigenvalues $n,~l,~(l-n)$, and $(l+n)$, respectively, which
reflects the fact that only two of them are independent.  In 
order to fix the coefficients in the nine expressions and 
extract the 27 simultaneous eigenstates of all operators in 
Table III, we require that these expressions be eigenstates 
of a third independent operator, say $XXX$.   We may write 
the 27 basis states in analogy with \Eq{row5} as
\be
   \ket{n,l,m}_{G} = {1 \over \sqrt{3}} 
  ~\sum_{k=0}^2 C_{nlmk} \ket{k,n+k,l+k}_{zzz}, 
\label{GHZ1}
\ee
where $|C_{nlmk}| = 1$, the subscript G identifies the 
entangled basis {\it set} G$(Z^2ZI,Z^2IZ,XXX)$, and 
$n,~l$ and $m$ are the eigenvalues of (the Hermitean 
counterparts of) the three operators listed, respectively.  
For example, 
$H(XXX) = [XXX - (XXX)^{\dag}]/i\sqrt{3} \rightarrow m$.

The presence of the operator $XXX$ in this list might suggest 
the alternative separable basis set S$(XXX)$ for the expansion 
of the $\ket{n,l,m}_{G}$ states.  However, such expansions 
would require sums of nine terms, not three.  The basis 
S$(ZZZ)$ is special in reducing the expansion to its 
``simplest'' form.   This situation is analogous to that of 
three-qubit GHZ states, where the``simplest'' form consists of 
two-term expansions in separable states.  There too the 
simplification depends on a special choice of quantization 
axes for the individual qubits, other choices requiring 
four- or eight-term expansions.   

The ``three-qutrit GHZ'' states, 
like their three-qubit counterparts, share the property that 
all one-particle reduced density matrices are proportional to 
the identity.  The proof of this fact again follows directly 
from the operators themselves:   We expand the projection 
operator $\PAa$ of any such basis state in terms of the MCS
operators using  Eq. \ref{INVS} of Appendix A.  Since no 
operator in the expansion has more than a single identity 
factor (apart from the $III/27$ term), the partial trace 
of $\PAa$ over any two qutrits is proportional to the 
identity on the other, for example:
\be
  \rho_1 = \Tr_{2,3} \PAa = {1 \over 27} \Tr_{2,3} I 
  = {1 \over 3} I_1.
\ee

There are subtle differences among 3-qutrit G bases analogous
to those occuring in 2-qutrit B bases, in which the expansion 
of \Eq{GHZ1} takes slightly different forms.  For example, 
suppose that all operators on the third qutrit are 
interchanged with their squares in Table III.  Then sums and
differences are interchanged wherever the third qutrit is 
involved, and the expansion in the same separable basis 
S$(ZZZ)$ takes the form
\be
   \ket{n,l,m}_{G'} = {1 \over \sqrt{3}} 
  ~\sum_{k=0}^2 D_{nlmk} \ket{k,n+k,-l-k}_{zzz}, 
\label{GHZ2}
\ee
where G$'$ refers to G$(Z^2ZI,Z^2IZ^2,XXX^2)$, and $m$ is the 
eigenvalue of $H(XXX^2)$, {\it etc}.  
Clearly there are four distinct expansions of the type seen 
in Eqs. \ref{GHZ1} and \ref{GHZ2}, corresponding to the four 
ways of distributing ($-$) signs among the $k's$.

The Aharonov state $\ket{{\cal A}}$ that is used in the
solution of the Byzantine agreement problem \cite{Fitzi} is
the (unique) spin singlet state of three qutrits.  It is the
superposition of all six permutations of 0, 1, and 2 among the
three qutrits, with coefficients ($1/\sqrt{6}$) for even 
permutations and ($-1/\sqrt{6}$) for odd.  Thus, it is the
superposition of two states of the form of \Eq{GHZ1},
\be
  \ket{{\cal A}} = {1 \over \sqrt{2}} \big(~\ket{1,2,1}_G -
  \ket{2,1,1}_G~ \big)
\label{Ah}
\ee
where the third entry in the kets is the eigenvalue of $XXX$,
which is a cyclic permutation operator, and the first two 
entries distinguish even and odd permutations.

Let us finally illustrate MCS types that produce basis sets of 
mixed entanglement, in which one particle is unentangled while 
the other two are maximally entangled in the Bell-like states 
of the previous section.  One such subset that singles out the 
first particle is generated by $ZII, IZX$, and $IYZ$.  The 
corresponding basis set can be denoted by 
SB$[Z_1;(ZX,YZ)_{2,3}]$, which indicates products of 
$Z$-states for particle 1 with Bell states for particles
2 and 3.  States within this basis set can be expanded using
the coefficients defined in \Eq{row5}:
\be
  \ket{n,l,m}_{SB} = {1 \over \sqrt{3}} 
  ~\sum_{k=0}^2 C_{lmk} \ket{n,k,l-k}_{zzx},
\label{PiB}
\ee
where the subscript SB is short for SB$[Z_1;(ZX,YZ)_{2,3}]$.   
Here the first index does not vary in the sum; in other basis 
sets such as SB$[Z_2;(ZX,YZ)_{1,3}]$ the second would be held 
constant.  A maximum of 12 such basis sets can be mutually 
unbiased:  Any qutrit can be singled out to be unentangled, 
and in each case four basis sets are possible ($V, X, Y$, or 
$Z$).  In a partition where this maximum is realized, the 
remaining 16 basis sets must all be of the GHZ (G) type.   

To understand the range of partitioning structures, we classify 
the 728 operators as one-body (two identity factors), two-body 
(one identity factor), and three-body (no identity factors).
The total number of each type is given in Table IV, and compared
with the numbers that occur within each type of MCS.  This table 
shows that there can never be more than 4 mutually unbiased 
separable bases, since this would require more than the existing
24 one-body operators.   Similarly, there can never by more than
12 partially entangled bases.  More generally, the numbers of 
each type that may coexist within any partitioning structure 
must satisfy the relations
\bea
  N(\hbox{G}) &=& 16 + 2 N(\hbox{S}) \nonumber \\
   N(\hbox{SB}) &=& 12 - 3 N(\hbox{S}), 
\label{coex}
\eea 
where $N(\hbox{S}) = 0,1,...4$ is the number of separable bases. 
The number of partially entangled (SB) bases may be 0 - 12 in 
steps of 3, and the number of GHZ bases may range from 16 - 24 
in steps of 2.  Equations \ref{coex} require that no other MCS 
types exist, and one can easily verify that this is indeed the 
case \cite{three}.   Similar coexistence conditions exist for
the three qu{\it bit} case, where analogous types of mutually
unbiased basis sets occur.  However, the new and surprising 
aspect of the present results is the existence of a 
{\it minimum} number of GHZ bases.

\begin{table}
$$
\vbox{
%  \halign{
% # \hfil \quad & \hfil # \hfil \quad & \hfil # \hfil \quad\cr
%\noalign{\hrule}
%\noalign{\smallskip}
%    2 qutrits    & 1-body & 2-body                 \cr
%\noalign{\hrule}
%\noalign{\smallskip}
%        all operators & 16 & 64                    \cr
%	one S basis & 4 & 4                         \cr
%	one SB basis & 0 & 8                        \cr
%\noalign{\smallskip}
%\noalign{\hrule}
%\noalign{\smallskip}
%}
  \halign{
 # \hfil \quad & \hfil # \hfil \quad & \hfil # \hfil \quad & 
                                       \hfil # \hfil \quad\cr
\noalign{\hrule}
\noalign{\medskip}
     operators   & 1-body & 2-body & 3-body            \cr
\noalign{\medskip}                   
\noalign{\hrule}
\noalign{\medskip}
        total numbers & 24 & 192 & 512                 \cr
	one S basis & 6 & 12 & 8                       \cr
	one SB basis & 2 & 8 & 16                      \cr
	one G basis   & 0 & 6 & 20                     \cr
\noalign{\medskip}
\noalign{\hrule}
}}  
$$                            
\caption{The distribution of operators within the three types
of maximally commuting subsets:  separable (S), partially
entangled (SB), and totally entangled (G).   These profiles
determine which partitioning structures are possible.} 
\end{table}

\section{Concluding Summary}

The foregoing three sections illustrate how factorizations of the 
Pauli Group relate to operator partitions, and hence to mutually 
unbiased basis sets.   The factor groups of order $3^N$, for
example ($\CE,{\cal R,L})^{\otimes N}$, are isomorphic to 
every maximally commuting subset (MCS) in
the partition.   Every MCS has $N$ generators, which provide the 
necessary quantum numbers to label the associated basis states by 
$N$-digit trinary numbers.   The Hermitean counterparts of these
generators form the ``complete sets of commuting observables.''  
The $3^N+1$ distinct sets are mutually unbiased:  Any state for 
which one set takes definite values produces perfectly random 
outcomes for all other sets.  This statement applies equally well
to the full MCS's.

New aspects emerge as one proceeds to larger numbers of qutrits.  
In the case of a single qutrit there is a unique partitioning of 
the eight operators into four MCS's.  

In the case of two qutrits, there are 48 distinct partitionings,
but they all have the same structure, producing four separable 
bases and six totally entangled (Bell-like) bases.  Each of the 
10 MCS's has two generators, so that all quantum numbers 
are two-digit trinary numbers.   Those associated with the 
Bell-like basis states refer exclusively to sums or differences 
of particular one-qutrit attributes, and in doing so they impose 
perfect randomness on {\it all} one-qutrit attributes. 

The three qutrit case exhibits distinct partitioning structures, 
allowing for the coexistence of three different types of basis 
sets according to Eqs. \ref{coex}:   The GHZ-like 
states are defined by three-digit trinary numbers, each digit 
referring to sums or differences of two or three one-qutrit 
attributes.  At least one digit must refer to all three qutrits,
as is seen from the generators.  All one-qutrit attributes are 
perfectly random in the GHZ-like states.

The latter two cases illustrate a sort of complementarity 
between individual and joint properties of a system of particles.
This complementarity is expressed through operator sets rather
than individual operators, and in general all $N$ generators 
are required :   With $N=3$, suppose we choose two generators, 
$Z^2ZI$ and $Z^2IZ$, from Table III.   A third choice $XXX$ 
produces a GHZ basis, while the alternative choice $ZII$ 
produces a separable basis.   In the latter case, $XXX$ is 
random while in the former, $ZII$ is random (since {\it every} 
operator outside the MCS of the generators is random.  In this 
sense the concept of unbiasedness of operator sets incorporates 
aspects of both complementarity and contextuality.  The 
compatibility of two operators does not rest on their 
commutativity alone, but on the choice of other operators 
to be measured.
\vskip.3truecm

\centerline{\bf Acknowledgements}
\vskip.3truecm

I would like to thank Jagdish and Suranjana Luthra for 
many stimulating discussions about the subject of this work.

\vskip.3truecm 

\centerline{\bf Appendix A}

\vskip.3truecm
In this appendix we review a general theorem proven in 
\Ref{Bandyo}:  A full complement of mutually unbiased basis sets 
exists if and only if a partitioning, complete and orthonormal
set of operators exists.   We follow the notation  of a similar 
proof given specifically for $N$-qubit systems in \Ref{LBZ}. 

First, suppose that a full complement of $d+1$ mutually unbiased 
basis sets exists in a Hilbert space of dimension $d$.  In terms 
of projection operators $\PAa = \ket{A,\alpha} \bra{A,\alpha}$, 
where $A = 1,...,d+1$ denotes the basis and $\alpha = 1,...,d$ 
the state within it, the orthonormality of each basis is 
expressed by
\be
% \PAa \PAb = \delta_{\alpha\beta} \PAa 
  \Tr (\PAa \PAb) = \delta_{\alpha\beta},
\label{PAPA}
\ee
and the unbiasedness of different bases ($A \neq B$) by
\be
  \Tr (\PAa \PBb) = d^{-1}.
\label{PAPB}
\ee
Corresponding to each basis set, we may {\it define} a maximally 
commuting set of unitary operators, $\UAa$, where $a = 1,...,d$, 
including the identity, $I \equiv U^A_d$, by their 
spectral representations
\be
  \UAa = \sum_{\alpha = 1}^d \epsaa \PAa.
\label{SPEC}
\ee
Unitarity of $\UAa$ requires that $|\epsaa| = 1$, and 
$U^A_d = I$ requires $\eps_{d\alpha} = 1$.   We further 
stipulate that the rows of the ``$\eps$ matrix'' be orthogonal, 
\be
   \sum_{\alpha = 1}^d \epsaa^* \epsba 
   \equiv (\eps_a,\eps_b) = d~\delta_{ab},
\label{EAEB}
\ee
{\it i.e.}, that the scaled matrix $\eps/\sqrt{d}$ be unitary,
\be
  d^{-1} \eps^{\dag} \eps = I.
\label{UNIT}
\ee
In consequence, the $\UAa$ operators form an orthonormal set 
(with the exception of the redundant identities $U_d^A$), 
that is,
\be
  \Tr (U_a^{A\dag}\UBb) = \sum_{\alpha,\beta} \epsaa^* 
  \epsbb \Tr (\PAa \PBb) = d~\delta_{AB}\delta_{ab},
\label{ortho}
\ee
where Eqs. \ref{PAPA} and \ref{EAEB} are used if $A=B$, and
Eqs. \ref{PAPB} and $\sum \epsaa = 0$ (for $a \neq d$) are used
otherwise.  Equation \ref{ortho} shows that the $d^2-1$ traceless 
$\UAa$ operators, together with the identity, form a complete and 
orthonormal set.   The partitioning property is guaranteed by 
Eq. \ref{SPEC}, which implies commutativity within subsets ($A$).

The converse of the above theorem may be demonstrated immediately.
Assume that a complete and orthonormal set of operators $\UAa$
exists with the partitioning property.   Append the identity to
each maximally commuting subset and apply the inverse of the
transformation defined by Eq. \ref{SPEC}, exploiting \Eq{UNIT}:
\be
  \PAa = d^{-1} \sum_{a=1}^d \epsinaa \UAa.
\label{INVS}
\ee
The required properties (Eqs. \ref{PAPA} and \ref{PAPB}) are 
easily verified.   Equation \ref{INVS} is useful in Sections
III and IV.

As a final note, the operators $\UAa$ need not be unitary.  One 
can choose the matrix $\eps$ to have real entries making $\UAa$
Hermitean, for example.  The matrix $\eps/\sqrt{d}$ must then be
orthogonal to ensure orthonormality of the set $\{\UAa\}$.  
\vskip.3truecm 
\centerline{\bf Appendix B}
\vskip.3truecm 
In this appendix we prove the relationship between partitionings 
of the operator set $\{\UAa\}$ and factorizations of the Pauli 
Group, $\{(1,\om,\omsq) \times \UAa\}$.  Simply stated, 
{\it Any maximally commuting subset (say $A$) in the partition 
corresponds to the identity element of a Pauli factor 
group.  The other elements contain one and only one operator 
from every other MCS ($B \neq A$).   The factor group is
isomorphic to every MCS (plus the identity, and in triplicate).} 

The following proof is given for $N$ qutrits (Hilbert space 
dimension $d = 3^N$), although clearly the theorem is more 
general.

Consider the complete orthonormal set of unitary operators 
$\{\UAa\}$ that partitions into maximally commuting subsets 
$A = 1,...,d+1$, where $a = 1,...,d-1$ denotes operators within 
each subset.   The Pauli Group consists of these operators, plus
the identity, all {\it in triplicate} ({\it i.e.}, multiplied by 
the three phase factors $\omn$), and is thus of order 
$3d^2 = 3 \cdot 9^N$.   A subgroup $\CF$ consists of any 
maximally commuting subset (say $A$), plus the identity, in 
triplicate:
\be
  \CF = (1,\om,\omsq) (U^A_1,...,U^A_d),
\label{CALE}
\ee
where $U^A_d \equiv I$.  The inclusion of the phase factors makes
$\CF$ an invariant subgroup which, consisting of $3d$ elements, 
defines a factor group of order $d$.   All of the factor group 
elements are generated by the multiplication 
\be
  \CFB = \UBb \CF, 
\label{CALF}
\ee
with fixed $B \neq A$ and $b=1,...,d$, the last entry repreducing
the identity element ${\cal F}_d=\CF$.  The product rule of these 
$\CFB$ is identical to that of the $\UBb$ themselves, modulo 
$\omn$,
\be
  \CFB \CF_{c} = \UBb \CF U^B_c \CF = \UBb U^B_c \CF,
\ee
showing that indeed they form a factor group as advertised. 

To prove the theorem stated above, notice that every  
$\CFB$ ($b \neq d$) contains a distinct operator in $B$, each 
in triplicate.   But there are $d$ choices of $B$ ($B \neq A$),
every one of which must generate the same factor group elements.
Therefore every $\CFB$ ($b \neq d$) must contain a distinct 
operator from every maximally commuting subset other than $A$,
each in triplicate.   This accounts for the full constituency 
($3d$) of each $\CFB$, so the theorem is proved.

\begin{thebibliography}{99}
\bibitem{three} It is instructive to show 
     that any possible generator set leads to one of the three 
     MCS types listed.
\end{thebibliography}
\end{document}